\documentclass[12pt,preprint]{aastex}

\def\Msol{\thinspace\hbox{$\hbox{M}_{\odot}$}}

\def\a4{\hsize 17.0cm \vsize 25.cm}

\shorttitle{superwinds \& supernebulae}
\shortauthors{Silich et al.}

\begin{document}

\title{On the extreme positive star-formation feedback  condition in SCUBA 
sources}

\author{Sergiy Silich\altaffilmark{1}, 
Guillermo  Tenorio-Tagle\altaffilmark{1,2}, 
Casiana Mu\~noz-Tu\~n\'on\altaffilmark{3},
Filiberto Hueyotl-Zahuantitla\altaffilmark{1},
Richard W\"unsch\altaffilmark{4}}
\and
\author{Jan Palou\v{s}\altaffilmark{4}}
\altaffiltext{1}{Instituto Nacional de Astrof\'\i sica Optica y
Electr\'onica, AP 51, 72000 Puebla, M\'exico; silich@inaoep.mx}
\altaffiltext{2}{Sackler Visiting Fellow at the Institute of Astronomy,
University of Cambridge, U.K.}
\altaffiltext{3}{Instituto de Astrof\'{\i}sica de Canarias, E 38200 La
Laguna, Tenerife, Spain}
\altaffiltext{4}{Astronomical Institute, Academy of Sciences of the Czech
Republic, Bo\v{c}n\'\i\ II 1401, 141 31 Prague, Czech Republic}

\begin{abstract}

\end{abstract}
We present a detailed study of the hydrodynamics of the matter reinserted by
massive stars via stellar winds and supernovae explosions in young 
assembling galaxies. We show that the interplay between 
the thermalization of the kinetic energy provided by massive stars, 
radiative cooling of the thermalized plasma and the gravitational pull of the 
host galaxy, lead to three different hydrodynamic regimes.
These are: 
a) The quasi-adiabatic supergalactic winds. b) The bimodal flows, with mass 
accumulation in the central zones and gas expulsion from the outer zones 
of the assembling galaxy. c) The gravitationally bound regime, for which all 
of the gas returned by massive stars remains bound to the host galaxy and 
is likely to be reprocessed into futher generations of stars. Which of the 
three possible 
solutions takes place, depends on the mass of the star forming region 
its mechanical luminosity (or star formation rate) and its size. The model 
predicts that massive 
assembling galaxies with large star formation rates similar to those detected 
in SCUBA sources ($\sim 1000$ M$_\odot$ yr$^{-1}$) are likely to evolve in a 
positive star-formation feedback condition, either in the bimodal, or in the 
gravitationally bound regime. This implies that star formation in these 
sources may have little impact on the intergalactic medium and result 
instead into a fast interstellar matter enrichment, as observed in high 
redshift quasars. 


\section{Introduction}

In Cosmology today the study of the star formation negative feedback is 
recognized as one of the central issues regarding galaxy formation  
\citep{1986ApJ...303...39D,1998MNRAS.298..399F,2002ApJ...574..590S,
2002ApJ...579..247F}. The large UV photon output from massive stars 
and their violently deposited mechanical energy, make them indeed major 
players in the dynamics of the ISM and key negative feedback agents 
able to limit and stop star formation defining the efficiency of the 
process
\citep[see][and references therein]{1988ARA&A..26..145T,1999ApJ...517..103E}.
However, as shown by 
\citet{2005ApJ...628L..13T,2007ApJ...658.1196T}
and \citet{2008ApJ...683..683W}
for the case of massive and compact super stellar 
clusters, this may not be the whole story
as the stellar feedback may, in extreme cases, become positive. This would 
allow gravity to win over thermal pressure. This may 
also be the case if one considers star formation in the submillimeter (SCUBA) 
galaxies - high redshift sources with the highest ($\sim 1000$~\Msol \, 
yr$^{-1}$) star formation rates (SFRs) so far known. These 
have been observed in the submillimetre continuum (emission from warm dust 
in the rest-frame far-IR/submillimetre wavelengths), in the CO line emission 
associated with the cold molecular gas and in the near-infrared integral 
field spectroscopy, which deals with the rest-frame optical emission lines 
associated with the photoionized gas 
\citep[see, for example, ][and references therein]{1998Natur.394..241H,
2005MNRAS.359.1165G,2006ApJ...640..228T,2006MNRAS.371..465S}.

The global properties of SCUBA galaxies include a typical dynamical masses of 
$5 \pm 3 \times 10^{11}$~\Msol, a gas to dynamical mass fraction,
$f_g = M_{gas} / M_{dyn} \sim 0.25 - 0.3$, and a radius of about 1 - 3~kpc,
parameters consistent with the expected properties of massive spheroids in 
the early  Universe 
\citep{2005MNRAS.359.1165G,2006MNRAS.371..465S,2006ApJ...640..228T,
2008ApJ...684L..21S}.
The long star formation duty-cycle with
a time-scale $\sim 100 - 300$~Myr and the inhomogeneous nature of SCUBA 
sources favor a continuous star formation scenario
\citep{2006MNRAS.371..465S,2008ApJ...680..246T}.
Here we show how massive and violent star-forming events driven by a high 
rate of star-formation lead to positive feedback. Looking at some extreme 
cases, we identify radiative cooling as the agent capable of downgrading the 
impact of the stellar energy deposition, leading  inevitably to an extreme 
positive star formation feedback condition which should play  a 
major role in galaxy formation. We also show that gravity is another major 
player. The gravitational pull of the galaxy also leads to a positive feedback 
condition, particularly in compact proto-galactic sources. The gravitational
pull then prevents the formation of a supergalactic wind, retaining the 
injected and ablated matter within the star forming region, favoring its
accumulation and conversion into further generations of stars. 

In section 2 we examine the physical implications of massive star formation 
rates and discuss the physical limits between the various possible 
hydrodynamic regimes. The implications of events with a high SFR 
and our conclusions are given in sections 3 and 4.

\section{Star formation under a large SFR}

If one scales the evolutionary synthesis models 
\citep[e.g.][]{1995ApJS...96....9L} for star 
clusters generated by a constant star formation rate to 
the values inferred from the SCUBA sources 
\citep[$\ge$ 100 M$_\odot$ yr$^{-1}$; e.g.][]
{2005MNRAS.359.1165G,2006MNRAS.371..465S,2006ApJ...640..228T}, 
one sees that as a result of the continuous death and 
creation of massive stars, the UV photon output will level off at 
$\sim 10^{55}$ ionizing photons s$^{-1}$ after 3 Myr of the evolution.
The mechanical energy deposited by the evolving stars ($L_{SF}$) through 
winds and supernovae (SNe) will also increase, although not so rapidly, 
to reach a constant value $\sim 2.5 \times 10^{43}$ erg s$^{-1}$ after 40 Myr
of evolution. Accordingly, the mass violently returned to the ISM by
stellar winds and supernovae will amount to $3 \times 10^7$ M$_\odot$ 
after 10 Myr, reaching almost 10$^9$ M$_\odot$ after 100 Myr of
evolution. The absolute  values of all the above mentioned variables 
ought to be linearly scaled   by more than an order of magnitude, at 
the given times, if instead of a SFR equal to 100  M$_\odot$ 
yr$^{-1}$, one assumes the even larger values inferred for the most 
powerful SCUBA sources ($\geq$ 1000 M$_\odot$ yr$^{-1}$).

At first glance, such an energy deposition and such a vast amount of matter so
violently injected, would unavoidably lead to extreme massive outflows into the
intergalactic medium  
\citep[see, for example,][and references therein]{1990ApJS...74..833H,
2000MNRAS.314..511S,2002ApJ...574..590S,2003ApJ...597..279T,
2005ARA&A..43..769V}.
Supergalactic winds are believed to result 
from the full thermalization of the kinetic energy of the ejecta, through 
multiple random collisions within the star-forming volume \citep[see][]{1985Natur.317...44C}. Thermalization generates the large overpressure that continuously
accelerates the deposited matter to finally blow it out of the star-forming 
volume, composing a stationary superwind with an adiabatic terminal speed 
$V_{A\infty} = (2 L_{SF}/\dot M_{SF})^{1/2}$; where $L_{SF}$ and $\dot M_{SF}$ 
are the mechanical energy and mass deposition rates provided by stellar
winds and supernovae explosions within the 
star-forming volume. For this to happen, the ejecta
has to reach an outward  velocity equal to the sound speed ($c = 
(\gamma P/\rho)^{1/2}$ $\propto T^{1/2}$) right at the star-forming boundary,
$R_{SF}$, to then fulfill the stationary condition in which the rate at which 
matter is deposited equals the rate at which it streams away from the 
star-forming region: $\dot M_{SF} = 4 \pi R_{SF}^2 \rho_{SF} c_{SF}$, where
$\rho_{SF}$ and $c_{SF}$ are the values of density and sound speed at the
surface of the star forming region. However, as shown in the present series 
of papers 
\citep[see][]{2003ApJ...590..791S,2004ApJ...610..226S,2005ApJ...628L..13T,
2007ApJ...658.1196T,2008ApJ...683..683W},
when dealing with the outflows 
generated by massive bursts of star formation, the impact of radiative 
cooling becomes a relevant property, as is gravity, able to hold a 
fraction of the deposited matter within the star cluster volume.

In the case of an instantaneous burst of star formation, stellar winds and 
supernovae are able to remove the matter left over from star formation out
of the star cluster volume in just a few Myr 
\citep{2006A&A...445L..23M,2006ApJ...643..186T}
and so the hydrodynamic solution
considers only the matter reinserted by the massive stars. In the continuous 
star formation scenario however, a gas reservoir out of which a constant SFR 
could be sustained is required. This implies that besides the mass returned by 
supernovae and stellar winds, $\dot M_{SF}$, the flow may hold 
additional matter. This 
results from the destruction and mass ablation from star forming clouds 
and can be normalized to the star formation rate within the star forming 
region: ${\dot M}_{ld} = \eta_{ld} SFR$ , where $\eta_{ld}$ is the mass 
loading coefficient. 
The total mass input rate into the flow is then:
\begin{equation}
\label{eq1}
{\dot M} = {\dot M}_{SF} +  {\dot M}_{ld} = 
 \left(\frac{2 L_0}{V^2_{A\infty}} + \eta_{ld} \times 
  1 {\rm M_{\odot} yr^{-1}} \right) \frac{SFR}{1 {\rm M_{\odot} yr^{-1}}} ,
\end{equation}
where $L_0$ is the normalization coefficient, which relates the mechanical 
energy output rate, $L_{SF}$, with the SFR:
\begin{equation}
\label{eq2}
L_{SF} = L_0 (SFR / 1 {\rm M_{\odot} yr^{-1}}) .
\end{equation}  
Hereafter we shall adopt $L_0 = 2.5 \times 10^{41}$~erg s$^{-1}$ and 
$V_{A\infty} = 2750$~km s$^{-1}$.
These values result from Starburst 99 synthetic models
for a continuous star formation mode with a Salpeter IMF and sources between 
0.1\Msol \, and 100\Msol, for ages $t \ge 40$~Myr 
\citep{1999ApJS..123....3L}.
Note, that mass loading changes the outflow terminal speed, which in this 
case is smaller than $V_{A\infty}$: 
\begin{equation}
\label{eq3}
V_{\infty} = \left(\frac{2 L_{SF}}{\dot M_{SF} + \dot M_{ld}}\right)^{1/2} =
             \frac{V_{A\infty}} 
             {\left(1 + \frac{1 {\rm M_{\odot} yr^{-1}} \eta_{ld}
              V^2_{A\infty}}{2 L_0}\right)^{1/2}} .
\end{equation}

For the calculations our semi-analytic radiative stationary wind code 
\citep{2008ApJ...686..172S} and the 1D hydrocode ZEUS3D \citep{1992ApJS...80..753S}
as modified by \citet{2007ApJ...658.1196T} and \citet{2008ApJ...683..683W} 
were used. All numerical calculations were performed with an open inner 
and outer boundary conditions.
The hydrodynamic equations here include the 
gravitational pull from the baryonic and dark matter, both assumed to 
be homogeneously distributed inside the star forming region. In our approach 
it is also assumed that the mass of the flow is negligible compared to the 
dynamical mass of the system, and thus the self-gravity of the re-inserted 
gas is not included in the calculations. For example, the mass of the flow
within the star forming region, $M_{flow}$, normalized to the dynamical
mass of the system is: $M_{flow}/M_{dyn} \approx 2 \times 10^{-5}$ 
and $M_{flow}/M_{dyn} \approx 8 \times 10^{-3}$
in case of model 1 with a low SFR and model 4 with a high SFR, respectively.  
To obtain the stationary hydrodynamic 
solution one has to know the mechanical energy and mass deposition rates, 
which are defined by equations (\ref{eq1}) and (\ref{eq2}), and the radius of 
the star forming region, $R_{SF}$. We use the equilibrium cooling function, 
$\Lambda(T,Z)$, tabulated by \citet{1995MNRAS.275..143P} and set the metallicity of the 
plasma to the solar value in all calculations. Our reference models
are presented in Table 1. Here the first column marks the model in our list,
the ablation coefficient, $\eta_{ld}$, is presented in column 2, columns 3, 4 
and 5 present the radius, dynamical mass of the star forming region and the 
star formation rate, respectively. Column 6 provides information regarding 
the resultant hydrodynamic regime. 
\begin{table}[htp]
\caption{\label{tab1} Reference models}
{\small
\begin{tabular}{c c c c c c}
\hline\hline 
Model  & $\eta_{ld}$ & Radius & Dynamical mass & SFR & Regime \\
       & & (kpc)  & ($10^{11}$ M$_{\odot}$) & (M$_{\odot}$ yr$^{-1}$) &  \\
\scriptsize{(1)} & \scriptsize{(2)} & \scriptsize{(3)} &\scriptsize{(4)} 
                 & \scriptsize{(5)} &  \scriptsize{(6)} \\
\hline
1  &  0.5 & 2.5 & 2 & 2    &  Superwind \\
2  &  0.5 & 1.65 & 2& 2    &  Superwind  \\ 
3  &  0.5 & 1.2  & 2& 2    &  Gravitationally bound \\ 
4  &  0.5 & 2.5 & 2 & 1200 &  Bimodal \\
5  &  0.5 & 1.7 & 2& 1200  &  Bimodal \\
6 &  0.5 & 1.2  & 2 &1200  &  Gravitationally bound \\
\hline\hline
\end{tabular}
     } 
\end{table}

\section{The hydrodynamic regimes}

There are three major hydrodynamic regimes that develop within 
galaxies undergoing a large SFR. Which of the three possible solutions takes
place, depends on the mass of the star forming region and its position in the 
mechanical luminosity or SFR versus size ($R_{SF}$) parameter space. Figure 1 
presents the threshold lines, which separate proto-galaxies evolving in 
different hydrodynamic regimes. The left panel presents threshold lines for 
proto-galaxies whose dynamical mass is 
equal to $2 \times 10^{11}$\Msol \, for different values of the ablation 
coefficient ($\eta_{ld} = 0.1, 05, 1.0$, dotted, solid  and dashed lines,
respectively). Below the threshold lines radiative cooling has a negligible 
effect on the flow and the reinserted matter ends up as a superwind.
Above these lines radiative cooling leads to a bimodal regime in which some
of the reinserted matter within the densest central regions loses its
pressure and is unable to participate in the galactic wind. Instead it 
accumulates there fueling further stellar generations
\citet{2005ApJ...628L..13T,2008ApJ...683..683W}
. For compact star forming regions, to the left of 
the vertical lines shown in Figure 1  ($R _{SF} < R_{crit}$), gravity inhibits 
the formation of a superwind, leading instead to matter accumulation and to 
further generations of star formation. In these cases the sound speed at
the surface of the star-forming region is smaller than 
$\sim (G M_{dyn}/2 R_{SF})^{1/2}$, which is one half of the escape speed from 
the proto-galaxy surface. 
The threshold lines for less ($4 \times 10^{10}$\Msol) and 
more ($4 \times 10^{11}$\Msol) massive galaxies with $\eta_{ld} = 0.5$ are 
presented in Figure 1, right hand panel.
\begin{figure}[htbp]
\plottwo{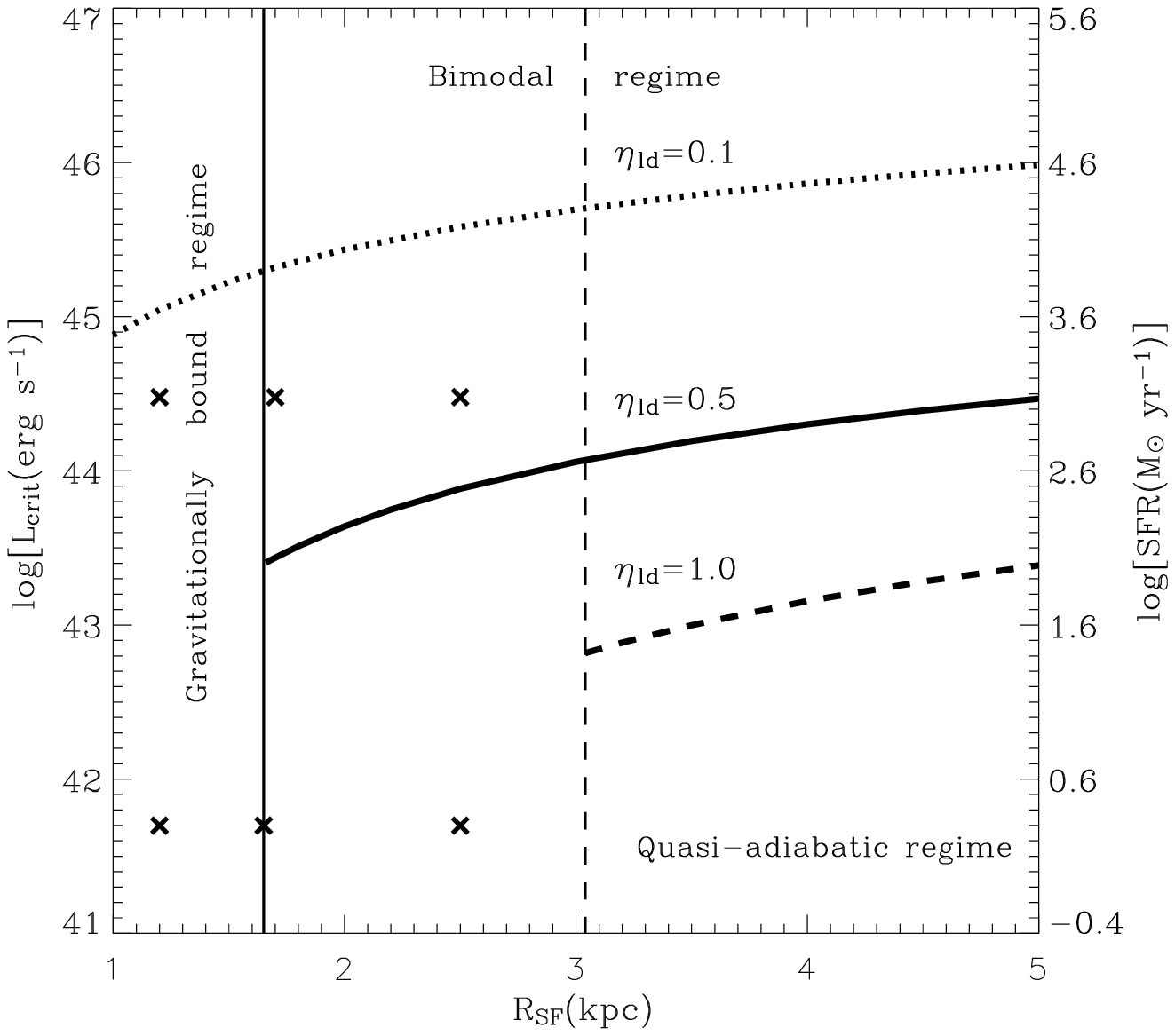}{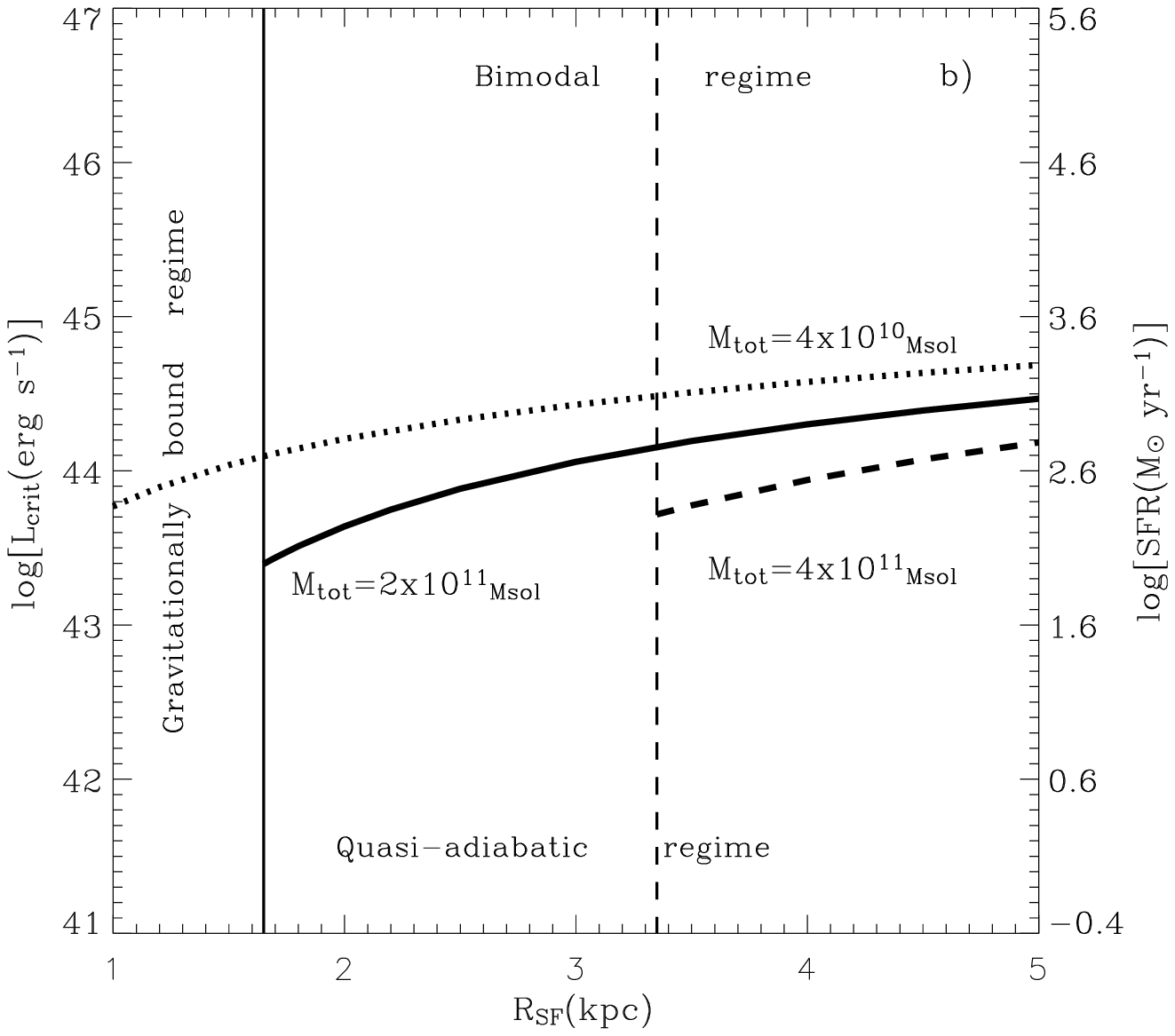}
\caption{The threshold SFR or energy input rate versus the threshold size. 
The left panel displays the threshold mechanical luminosity, SFR and critical 
radii for different values of $\eta_{ld}$ in the case when the dynamical 
mass in the star forming system equals to $M_{dyn} = 2 \times 10^{11}$\Msol.
The limiting energy input 
rate and its corresponding constant star formation rate (right-hand axis) 
above which strong radiative cooling inhibits the stationary superwind 
solution, as a function of the size of the star-forming region and the mass 
ablation coefficient, $\eta_{ld}$. These are terminated at the vertical
lines, which display the critical radii, $R_{crit}$. Gravity inhibits 
the formation of supergalactic winds in systems with smaller radii.
The right hand panel 
shows how the location of the threshold lines depends on the total mass 
of the star forming region. Several of the cases here presented are marked by 
crosses in the left panel. Note that the vertical line that marks the
critical radius ($R_{crit}$) for $\eta_{ld} = 0.1$ in the left panel, and for
$M_{dyn} = 4 \times 10^{10}$\Msol , in the right panel, lies within
1 kpc and thus is not shown.}
\label{fig1}
\end{figure}

As initially expected (see section 2) there is a large fraction of the 
parameter space that leads to stationary supersonic winds. In these cases, 
all the deposited matter, as well as that ablated from clouds, is able to 
escape from the gravitational well of the galaxy. For this to happen the flow 
has its stagnation point (the point where the velocity of the flow equals to 
0 km s$^{-1}$) right at the center of the galaxy ($R_{st} = 0$~pc) and its 
sonic point at the surface. The matter accelerates then through pressure 
gradients to reach supersonic velocities and form a supergalactic wind as it 
streams away from the galaxy.

Models 1 and 2 in Table 1 undergo such supergalactic winds. The distribution 
of the hydrodynamical variables in this case is shown in Figure 2. Here panel 
(a) presents the flow velocity (solid line) in the case of model 1 and 
compares this to the local sound speed, $(\gamma P(r)/\rho(r))^{1/2}$ 
and $V_{esc}(r) = (2 G M(r) / r)^{1/2}$, which is the escape velocity
at any distance $r$ outside the star-forming region.
The outflow velocity reaches the local sound speed value
right at the surface of the star forming region, then it accelerates rapidly 
to reach its terminal value of $\sim 740$ km s$^{-1}$ at a distance about 
4~kpc from the galaxy center. At this distance it already exceeds
the escape velocity and thus composes a supergalactic wind. Panels (b) and (c)
present the distributions of temperature and density in the flow.
The temperature drops from $\sim 2 \times 10^7$~K inside of the star forming 
region to $\sim 2 \times 10^6$~K at a 10~kpc distance from the galaxy whereas
the density drops from $\sim 4 \times 10^{-3}$~cm$^{-3}$ to less then 
$10^{-4}$~cm$^{-3}$ value. Such proto-galactic winds should be detected as 
sources of a diffuse X-ray emission, as it is the case in the 
local Universe 
\citep[e.g.][]{1992ApJ...397L..39C,2000MNRAS.314..511S,2005ApJ...635.1116S,
2009ApJ...697.2030S}:
\begin{equation}
      \label{eq.x}
L_X = 4 \pi \int_{R_{st}}^{R_{out}} r^2 n^2 \Lambda_X(T,Z) {\rm d}r ,
\end{equation}
where $n(r)$ is the atomic number density, $\Lambda_X(Z,T)$ is the X-ray 
emissivity \citep[see][]{2000MNRAS.314..511S}, $R_{out}$ marks the distance 
at which the calculations where stopped, usually set to 10 kpc. We set 
the lower integral limit to $R_{st}$ assuming that the X-ray emission 
interior to it is completely absorbed by the accumulated gas. 
The model predicts a growth in the X-ray
luminosity in the range from 0.3 keV - 8.0 keV as one considers larger SFRs. It is $L_x \approx 4 
\times 10^{-4} L_{SF} \approx 2 \times 10^{38}$erg s$^{-1}$ and reaches 
$L_x \approx 0.1 L_{SF} \approx 3 \times 10^{43}$erg s$^{-1}$  in the case of 
model 1 with a low SFR and model 4 with a high SFR, respectively.
Note that the X-ray emission is concentrated towards the star forming region, where the density of the
X-ray plasma reaches its maximum value and that in proto-galaxies with 
a high SFR a significant fraction of this emission may be absorbed by 
numerous dense proto-stellar clouds.
\begin{figure}[htbp]
\vspace{16.0cm}
\includegraphics{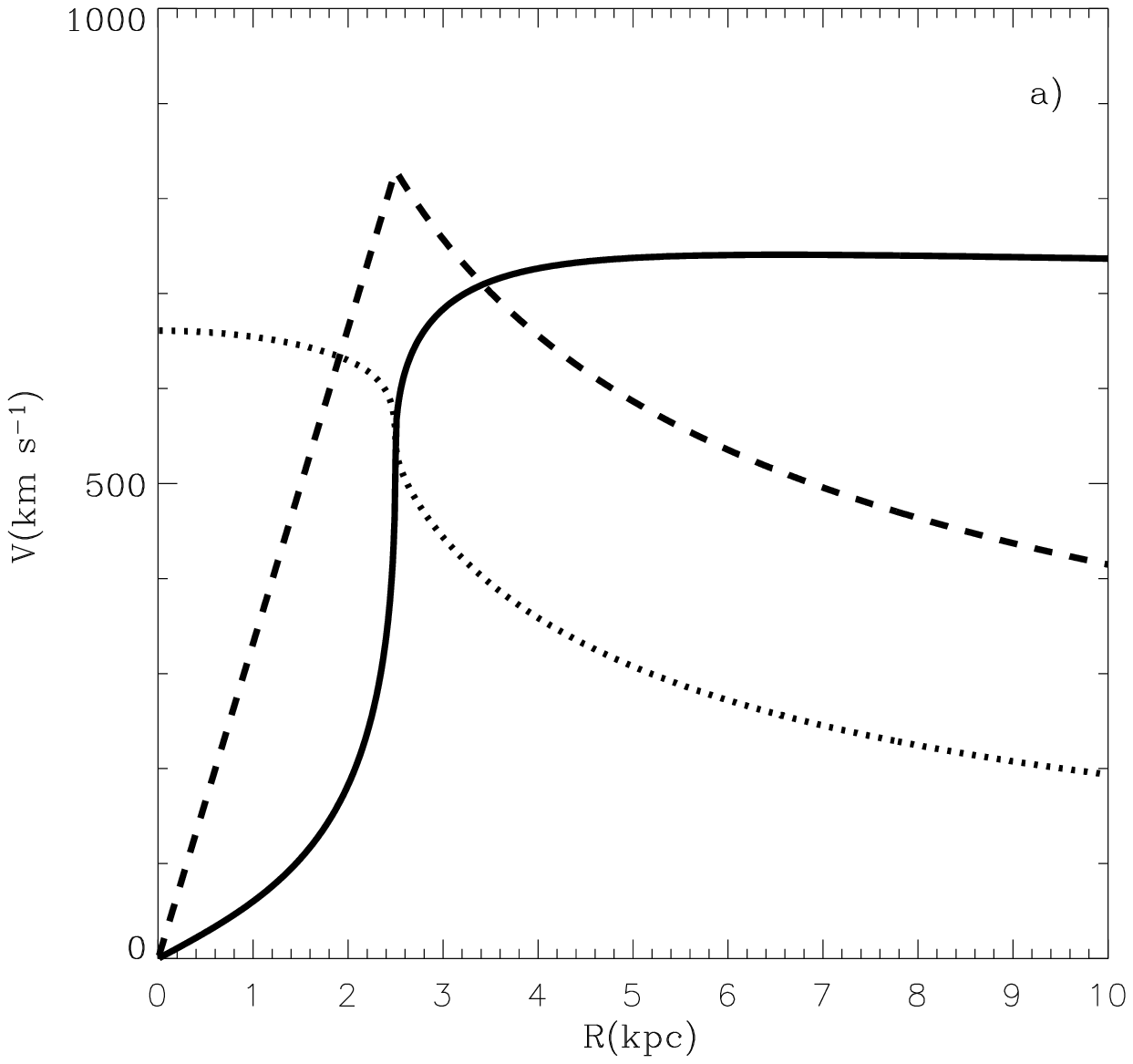}
\includegraphics{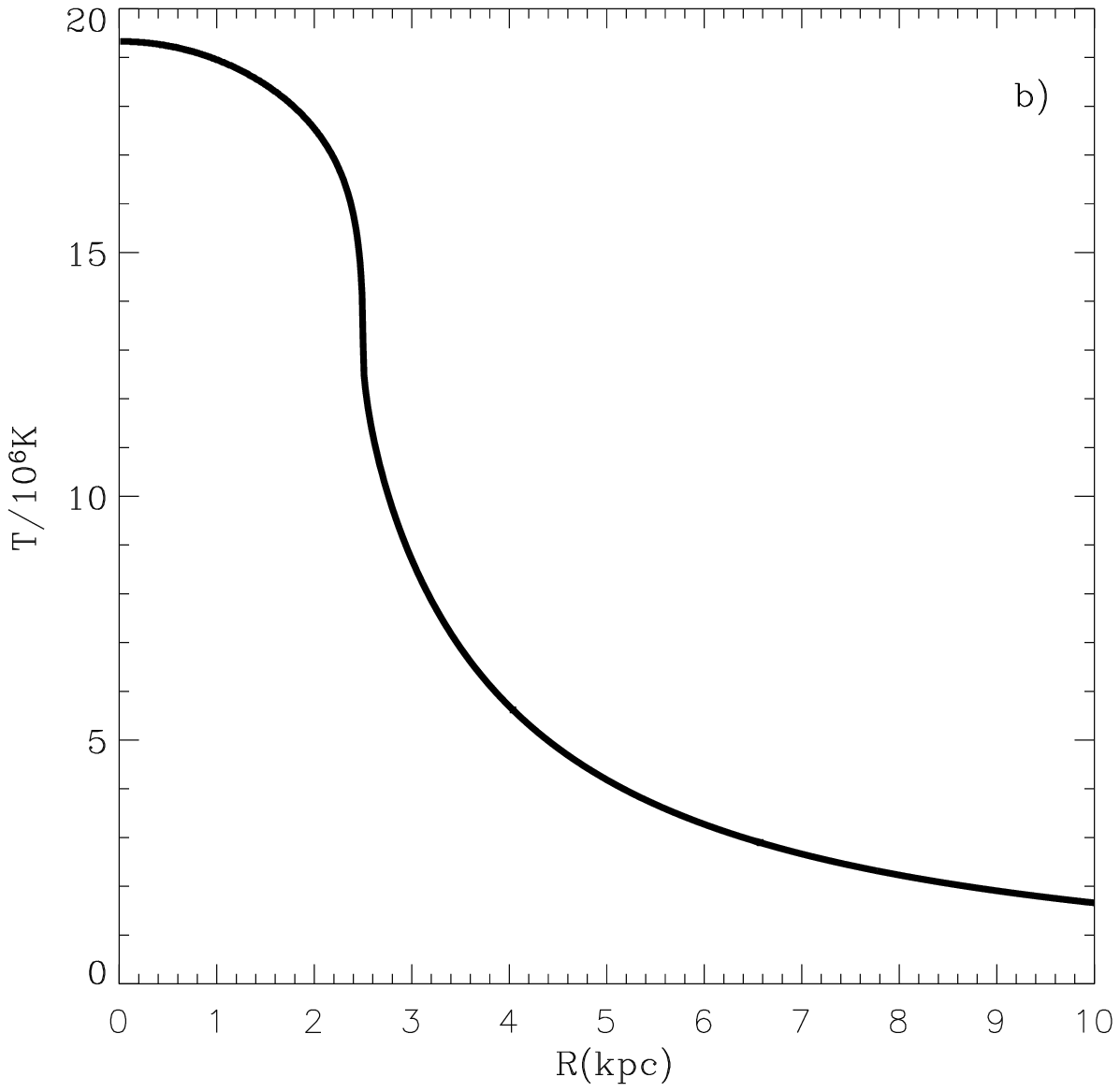}
\includegraphics{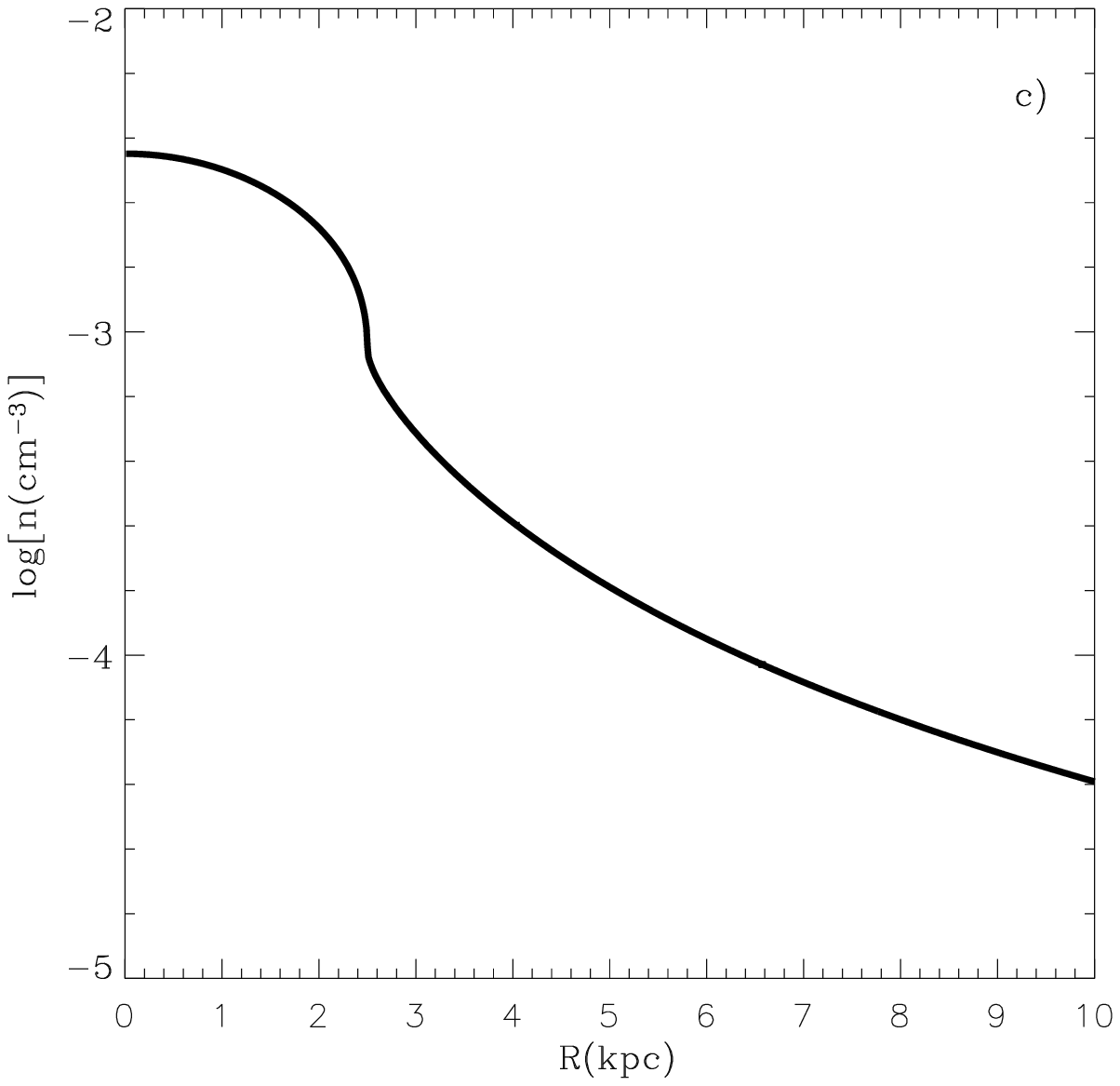}
\includegraphics{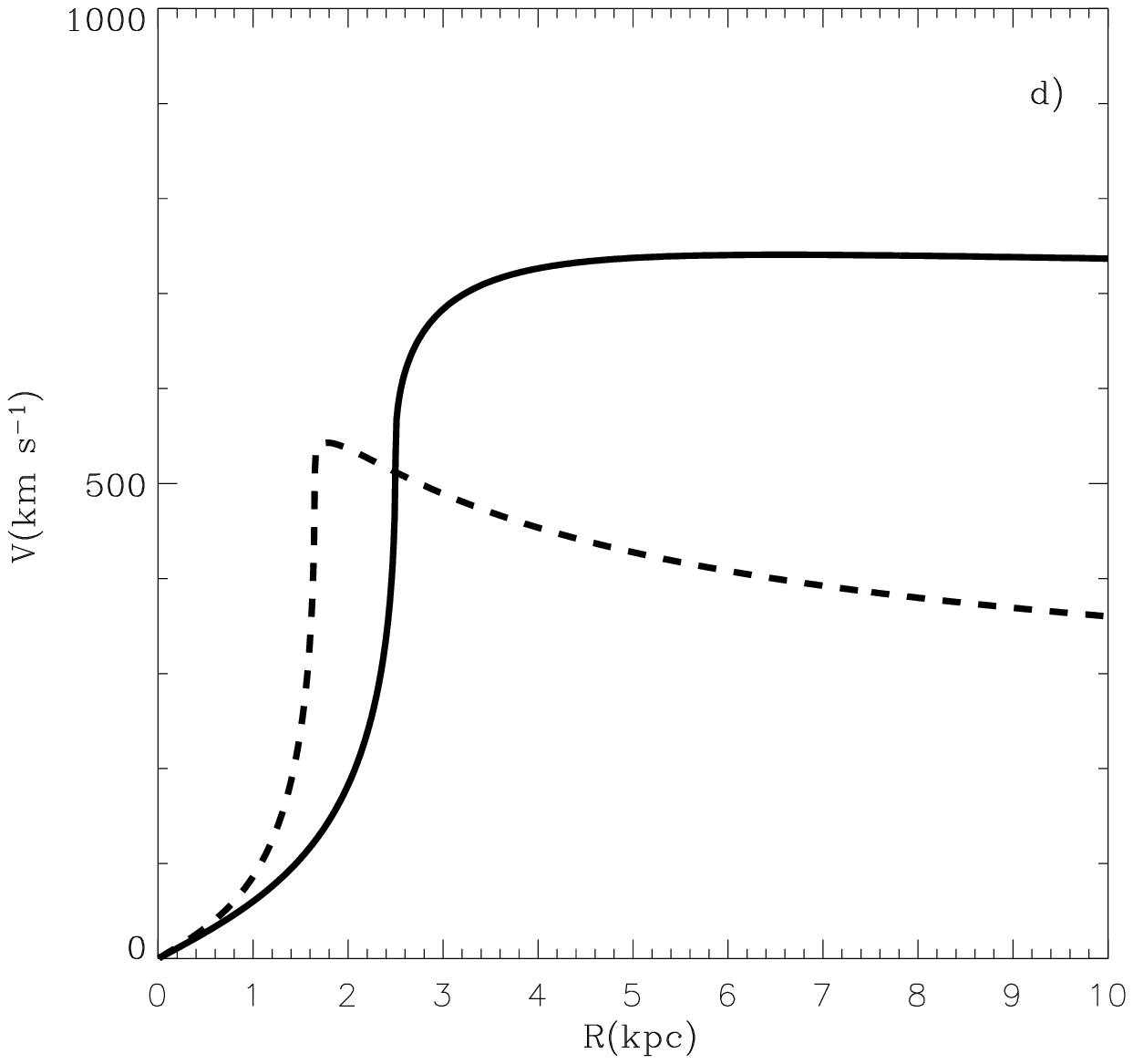}
\caption{The distribution of hydrodynamical variables in the supergalactic 
wind outflow. The calculations were provided for a proto-galaxy with a 
dynamical mass of $M_{dyn} = 2 \times 10^{11}$~\Msol, $R_{SF} = 2.5$~kpc,
SFR = 2 M$_{\odot}$ yr$^{-1}$ 
and $\eta_{ld} = 0.5$ (model 1). Panels a, b and c present the run of
velocity, temperature and particle number density, respectively. The dotted
and dashed line in panel a display the local sound speed and the value of 
$V_{esc}(r)$, respectively. Panel (d) compares the velocity distribution
in two proto-galactic winds emerging from sources of different size
(model 1 and 2, solid and dashed lines, respectively).}
\label{fig2}
\end{figure}

However, in the case of SCUBA sources gravity may affect the outflow 
significantly. Indeed, the escape speed at the surface of scuba sources, 
$V_{esc} = (2 G M_{dyn} / R_{SF})^{1/2}$, may reach $\sim 1000$~km s$^{-1}$, 
value which is approximately 10 times larger than in the case of young 
stellar clusters. In many cases it is larger than the sound speed in the 
thermalized plasma, and thus larger than the outflow velocity at the surface 
of the proto-galactic cloud.  The larger impact of gravity on the flow 
for progressively more compact systems with the same mass is shown in 
Figure 2, panel (d), which compares the run of velocity for models 1 and 
2 (solid and dashed lines, respectively). The maximum velocity is much 
smaller and the flow velocity drops significantly with distance to the 
proto-galactic cloud in the case of more compact star forming region  
(model 2, dashed line). Nevertheless, it ends up exceeding  the escape 
velocity value at a larger distance from the proto-galaxy center, forming a 
supergalactic wind. 
\begin{figure}[htbp]
\plottwo{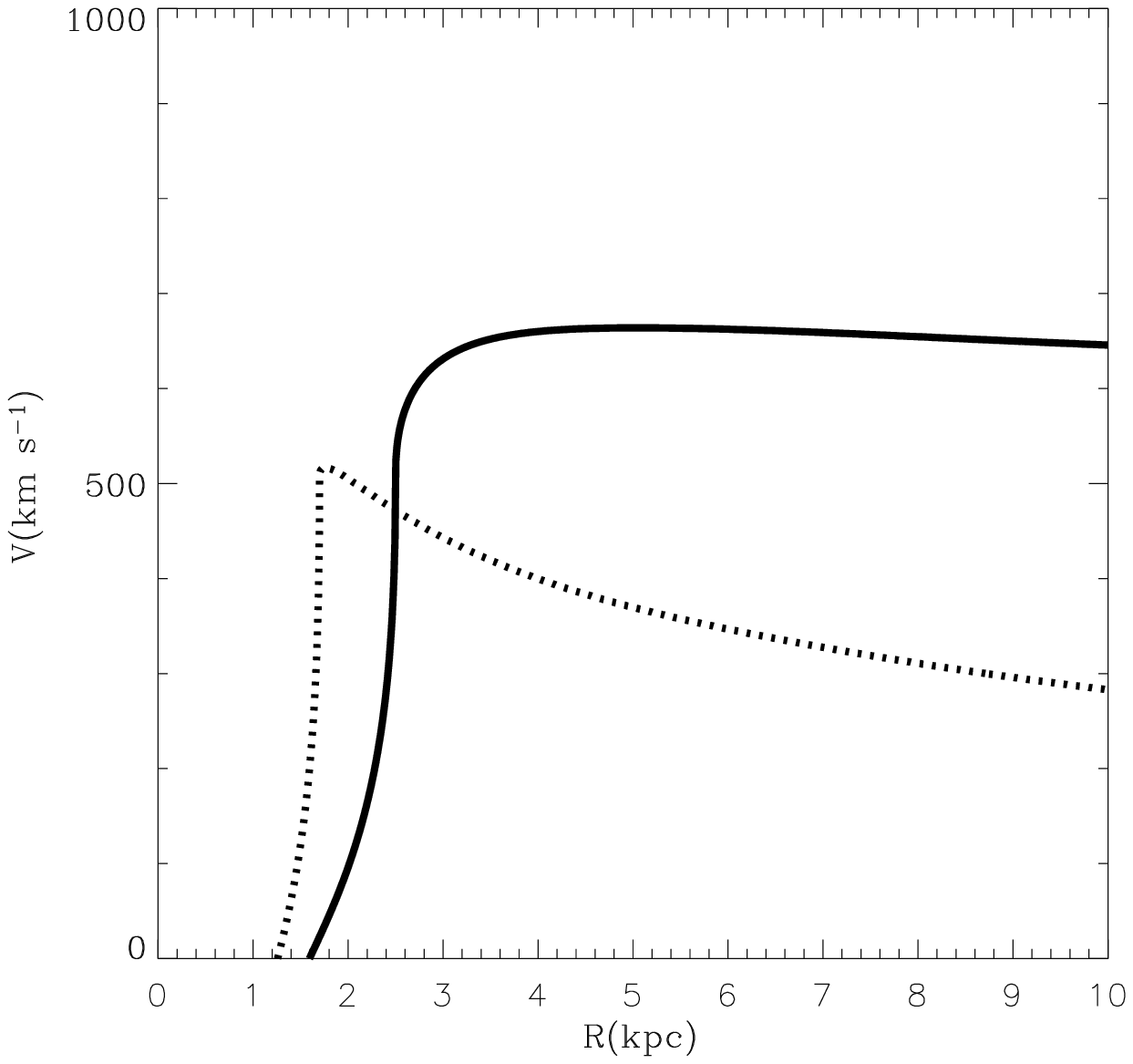}{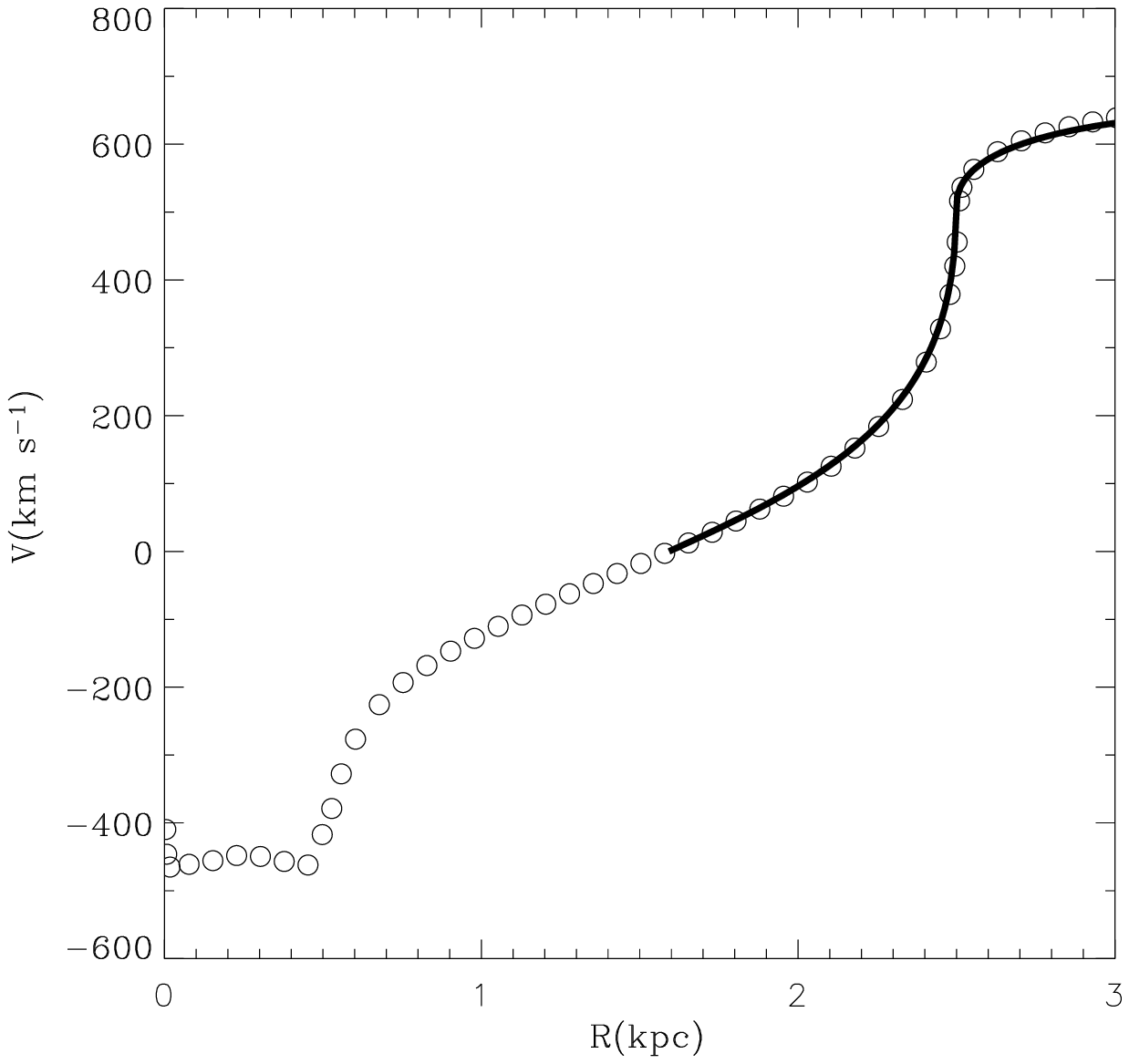}
\caption{The velocity of the flow for proto-galactic sources evolving
in the bimodal hydrodynamic regime. The calculations were carried out  for a 
proto-galactic cloud with a dynamical mass of $M_{dyn} = 2 \times 
10^{11}$~\Msol, SFR = 1200~\Msol \, yr$^{-1}$ and $\eta_{ld} = 0.5$. 
The left panel presents the results of the semi-analytic calculations for 
proto-galaxies with $R_{SF} = 2.5$~kpc and  $R_{SF} = 1.7$~kpc - solid and
dotted lines, respectively. In the right panel, the semi-analytic velocity 
distribution  for model 4 (solid line)  is  
compared with the numerical results (open circles).} 
\label{fig3}
\end{figure}

Proto-galaxies which lie above the threshold line (models 4 and 5) radiate a 
large fraction of the energy input rate within the star-forming volume, what 
leads to a bimodal hydrodynamic solution 
\citep{2007ApJ...658.1196T,2008ApJ...683..683W}. 
In this case radiative cooling depletes rapidly 
the thermal energy (and pressure) of the thermalized plasma in the 
densest central regions of the assembled galaxy, inhibiting the 
fast acceleration required to reach the sufficient speed to leave the star 
forming region. This prompts the stagnation radius, $R_{st}$, to move out 
of the starburst center as it is shown in Figure 3, left panel, 
where solid and dotted lines display the semi-analytic results for models 4 
and 5, respectively. Inside the stagnation radius the stationary 
solution does not exist and thus one cannot apply the semi-analytic model to 
this region 
\citep{2007ApJ...658.1196T,2008ApJ...683..683W}. However,
in Figure 3, right panel, we show the 
semi-analytic velocity distribution (solid line) in the case of model 4 and
compare this with the full numerical simulations (open circles) carried out 
with our Eulerian hydrodynamic code. The code includes also the gravitational 
pull from the matter located inside the
star forming region and a modified cooling routine, which allows for extremely 
fast cooling regimes \citep{2007ApJ...658.1196T,2008ApJ...683..683W}.
The numerical simulations show an  excellent agreement with our semi-analytic 
results. Both the positions of the stagnation point and the velocity 
profiles outside of the stagnation radius obtained with the semi-analytic and 
1D simulations are in good agreement. Thus above the threshold line the matter 
injected by massive stars and ablated from proto-stellar clouds inside the 
stagnation volume remains bound and is re-processed into new generations of 
stars despite the large amount of energy supplied by stellar winds and 
supernovae explosions. At the same time, the matter deposited by massive 
stars outside of this volume flows away from the star forming region as a 
supersonic wind. 

The impact of gravity becomes a crucial issue if the radius of the 
proto-galaxy is smaller or equal to the critical value, presented
in Figure 1 by vertical lines for different values of $\eta_{ld}$ and 
$M_{dyn}$. This occurs when the sound speed at the surface of the 
proto-galactic 
cloud becomes smaller than one half of the escape velocity, and the nominator 
in the momentum equation \citep[equation 12 in][]{2008ApJ...686..172S} goes to zero at 
the surface of the star forming region. In this case the flow velocity cannot 
exceed the sound speed value, the stationary solution vanishes and the 
proto-galaxy does not form a supergalactic wind. This regime is illustrated in 
Figure 4, which presents the results of full numerical simulations for 
a proto-galactic cloud with $R_{SF} < R_{crit}$ (model 6). Here the
quasi-adiabatic wind solution for a proto-galaxy with $M_{dyn} = 2 
\times 10^{11}$\Msol , $R_{SF} = 2.5$~kpc, SFR = 40~\Msol \, yr$^{-1}$ and 
$\eta_{ld} = 0.5$ was used as the initial condition for simulations. 
However, the time evolution was followed assuming the input parameters of
model 6 (see Table 1). The 
initial wind solution transforms rapidly into a complex flow with a number of 
discontinuities and negative velocities inside the star-forming region. 
Note that the stagnation radius is in this case at $\sim 1$ kpc.
However the matter deposited between this radius and the edge of the 
star-forming region is unable to produce a superwind and instead it cools 
down and ends up falling towards the center. Our open boundary condition
does not allow for the accumulation of this gas and that leads after a 
readjustment period to a recurrent cycle in which some fraction of the 
deposited matter first flows away but then cools down and falls back 
towards of the star-forming region. This causes the compression and 
storage of the hot gas into a dense shell, which is driven inwards by gravity.
The supersonic encounter of the outer gas with the dense shell
results into the formation of a shock wave. This at later times ($t >
30$Myr, dotted lines in Figure 4) produces the parcel of hot gas 
infalling behind the cold shell. The shell drives at all times a sound wave
into the hot inner zones, what results into noticeable enhancement of 
temperature and speeds up the infalling gas ahead of the shell as it is 
displayed by the dotted line in Figure 4.  
The simulation ended up at $\sim 30$~Myrs when all matter 
located inside a computational domain is falling towards the center of
the proto-galactic cloud. 

Thus compact proto-galaxies with $R_{SF} < R_{crit}$ trap the injected 
matter and are not able to form superwinds regardless of their energy 
output or SFR. 
\begin{figure}[htbp]
\plotone{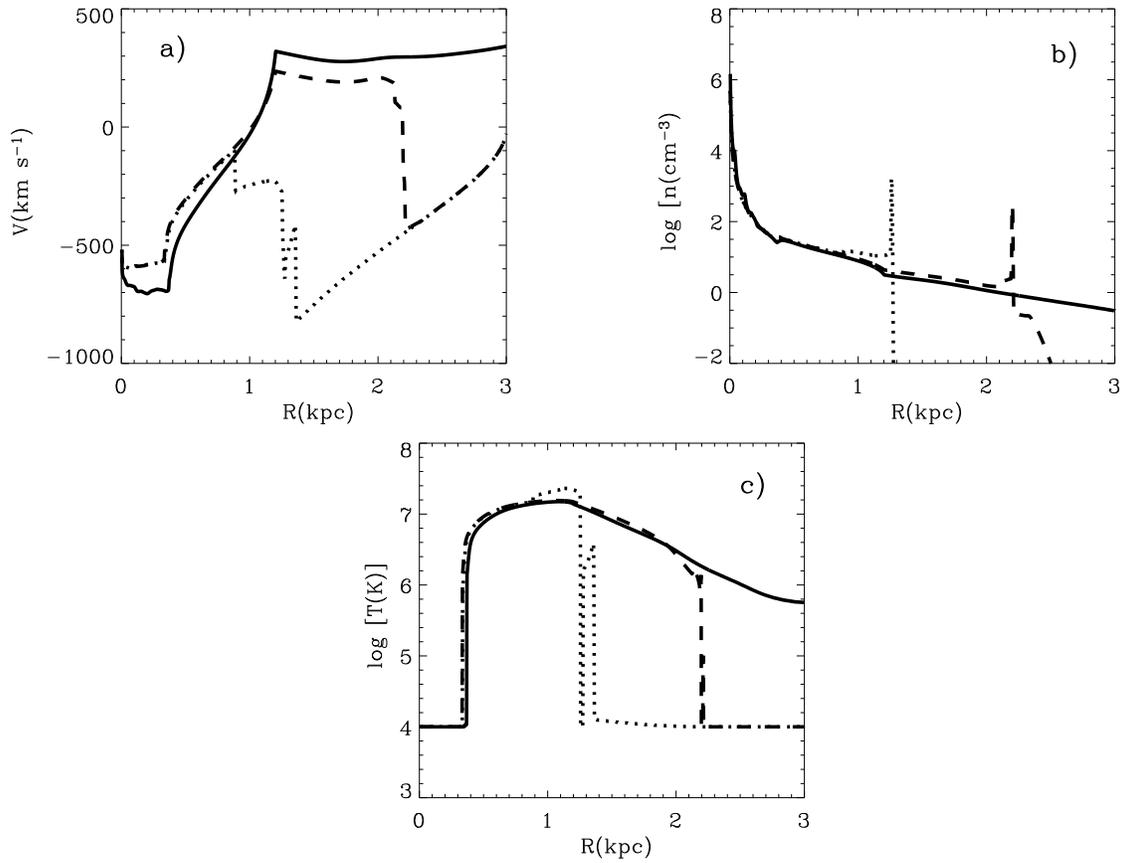}
\caption{The gravitationally bound regime. Panels a, b and c present 
runs of velocity, density and temperature at t = 22.2~Myr, 29.5~Myr
and 31.4~Myr (solid, dashed and dotted lines, respectively) in the
case 6 from Table 1.}
\label{fig4}
\end{figure}
A number of semi-analytic calculations have led us to infer that below the 
mechanical luminosity threshold line, $R_{crit}$ becomes slightly smaller for 
proto-galaxies with a 
given $M_{dyn}$ and $\eta_{ld}$ and a decreasing SFR. However the calculations
showed that the difference in the value of the critical radius usually does 
not exceed $\sim 100$~pc. Therefore we have adopted the value of $R_{crit}$, 
calculated for a proto-galaxy with the threshold SFR as the gravitationally 
bound limit for all galaxies with the same $M_{dyn}$ and $\eta_{ld}$. The 
adopted critical radii, $R_{crit}$, for galaxies with various $M_{dyn}$ and 
$\eta_{ld}$ are displayed in Figure 1 by thin vertical lines. 
Figure 5 shows how the critical radius depends on $\eta_{ld}$ and 
the dynamical mass of the system.
\begin{figure}[htbp]
\plotone{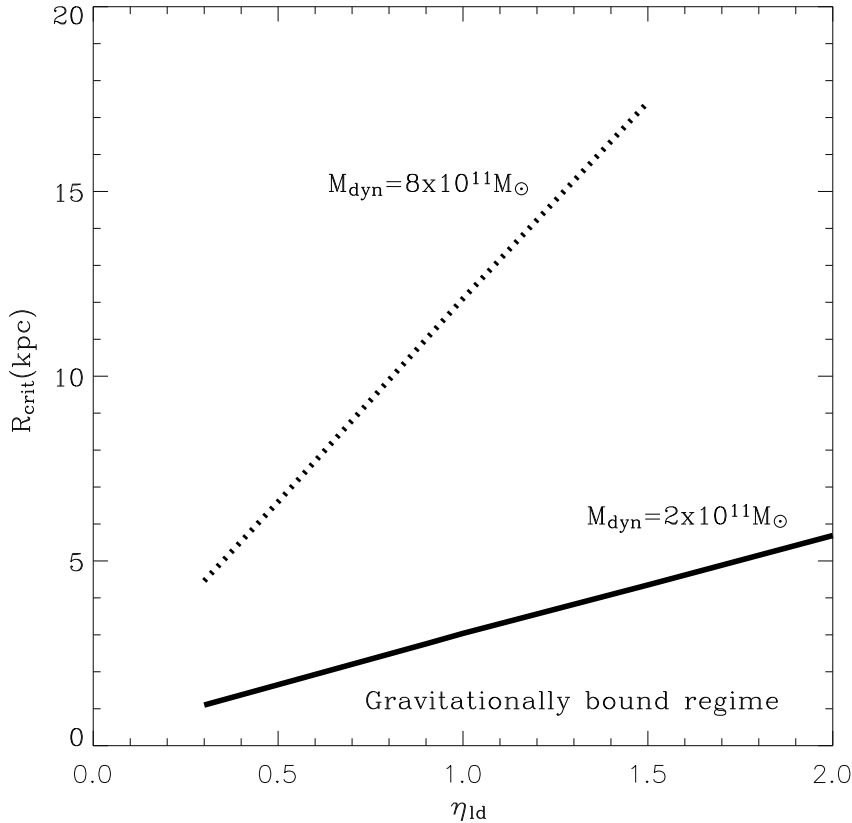}
\caption{The critical size ($R_{crit}$) of star-forming regions as a function
of the ablation parameter $\eta_{ld}$ and the dynamical mass of the 
proto-galaxy. The superwind feedback mode is inhibited in proto-galaxies 
with $R_{SF} < R_{crit}$. The matter returned by massive stars and ablated
from star-forming regions remains buried inside the star forming volume and 
as it accumulates it should lead to further stellar generations. 
The calculations were provided for proto-galactic clouds with 
$0.3 \le \eta_{ld} le 2$ and dynamical masses of $M_{dyn} = 2 \times 
10^{11}$~\Msol \, (solid) and  $M_{dyn} = 8 \times 10^{11}$~\Msol \, 
(dotted) lines, respectively.}
\label{fig5}
\end{figure}

Note that in very high redshift systems (i.e. just forming galaxies) the 
metallicity would be 
extremely low and hence radiative cooling will be substantially lowered. This could have potentially 
a large effect on the dynamics of the reinserted matter, favoring outflows. 
While this may be true for the first galaxies, 
observationally it is known that the metallicity of high redshift systems can be 
large in some cases, reaching values of several times solar. 
If metallicity grows rapidly as the galaxy forms, then radiative cooling will be even 
stronger than that predicted in our calculations further favoring the retention of 
the reinserted material. The threshold SFR line in Figure 1 moves up
a factor of 1.5 approximately in the case of the first galaxies with a 
metallicity $Z = 0.1Z_{\odot}$ and approximately 4.5 times down for older 
systems with a super solar abundance ($Z = 10Z_{\odot}$). Note that the
critical radii, $R_{crit}$, remain almost identical as one considers 
different metallicities. A time dependent 
solution accounting for the rapid change in the 
metallicity of the reinserted matter will be the subject of a forthcoming
communication.
 
\section{Discussion}

We have shown here that the thermalization of the kinetic energy provided
by vigorous star formation in young forming galaxies may lead to three
different hydrodynamic regimes, depending on the rate of star formation,
the proto-galaxy total mass and radius, and the rate of mass loading from
proto-stellar clouds. Large galaxies with low SFRs and small ablation
coefficient $\eta_{ld}$ form supersonic winds which carry from the star 
formation regions the matter returned by massive stars and that ablated
from proto-stellar clouds. Similar galaxies located in the SFR - $R_{SF}$ 
parameter space above the threshold line lose via a superwind a fraction of 
the deposited matter. The matter deposited by massive stars and that ablated 
from star forming regions in the inner zones of such galaxies 
becomes thermally unstable due to strong radiative cooling, accumulates and is 
to be re-processed there into secondary star formation. Finally, the thermal 
pressure in 
compact sources with radii $R_{SF} \le R_{crit}$ is unable to withstand
the gravitational pull of the galaxy. In such cases proto-galaxies 
retain all the  reinserted and ablated matter within the proto-galaxy volume 
and do not form supergalactic winds. 

The value of the ablation parameter $\eta_{ld}$ remains free in the theory.
However, one can get an idea about which values of $\eta_{ld}$ are reasonable 
considering sources without a secondary star formation, which evolve in the 
superwind regime. Then one can notice that in the case of star formation 
with a constant SFR, which terminates when the initial gas reservoir is 
completely exhausted, the global star formation efficiency, $\epsilon_*$, 
defined as the ratio of the stellar mass, $M_*$, to the initial mass, 
$M_{PG}$, of the proto-galactic cloud at this moment will be: 
\begin{equation}
\label{eq4}
\epsilon_* = \left(SFR - \frac{2 L_0}{V^2_{A\infty}} 
      \frac{SFR}{{\rm 1\Msol \, yr^{-1}}}\right) 
      \left(SFR + \eta_{ld} SFR\right)^{-1} = 
      \frac{0.9}{1 + \eta_{ld}} . 
\end{equation}
In equation (\ref{eq4}) the stellar mass, $M_*$, was calculated as the 
difference between the mass of stars formed during the evolutionary time 
$t$ and that reinserted by supernovae explosions and stellar winds:
\begin{equation}
\label{eq4a}
M_* = (SFR - {\dot M}_{SF}) \times t = \left(SFR - 
       \frac{2 L_0}{V^2_{A\infty}}\frac{SFR}{{\rm 1 M_{\odot} yr^{-1}}}\right) 
       \times t , 
\end{equation}
and the initial mass of the system, $M_{PG}$, is:
\begin{equation}
\label{eq4b}
M_{PG} = (1 + \eta_{ld}) \times SFR \times t .
\end{equation}
The star formation efficiency would be then $\epsilon_* = 90$\% if 
$\eta_{ld} = 0$ 
and approaches $\approx 30$\% value required to form a gravitationally
bound system \citep[e.g.][]{2001MNRAS.323..988G}
when $\eta_{ld} = 2$. Note, that the required star formation
efficiency may be smaller, and thus the upper limit for $\eta_{ld}$ larger,
if one considers a slow expulsion of the injected gas from the system
\citep{2007MNRAS.380.1589B}.

The predictions are thus that massive star-forming proto-galaxies with large
star formation rates similar to those detected in SCUBA sources 
($\ge 10^3$ M$_\odot$ yr$^{-1}$) evolve in a positive star-formation 
feedback conditions: either in the bimodal, or in the gravitationally
bound regime. Only proto-galaxies evolving in the bimodal 
regime will form supergalactic winds as it is in the case of
submillimeter galaxies SMM J14011+0252 \citep{2007ApJ...657..725N} and, probably,
SMM J221726+0013 \citep{2004MNRAS.351...63B}.   
Inevitably then, matter accumulation would follow in
the central zones or in the whole proto-galactic volume. Radiative 
cooling would then reduce the injected gas temperature, what would promote
an even stronger cooling and recombination, making the accumulated 
gas an easy target of the UV radiation field. Photoionization of this gas is 
to set an equilibrium temperature ($T_{HII} \le 10^4$~K), causing it to 
become Jeans unstable, leading unavoidably to its  collapse and  to 
the formation of new stars. Many stellar generations are expected in this 
scenario, until most of the mass, through its continuous recycling,  
has been converted into low mass stars with $M \leq 7$ M$_\odot$.
The resultant stellar populations and the ISM would then show a large 
metallicity spread.

Consequently, if the formation of large stellar spheroids (galaxy bulges or 
elliptical galaxies) occurs through a process of rapid matter accumulation and
further conversion of this matter into stars \citep[e.g.][]{2006MNRAS.371..465S}, which
would imply a large star formation rate, the expectations are thus that 
little or none of the returned matter, through winds and SN explosions, is 
going to be ejected out of the system. Instead it is to be reprocessed into 
further episodes of stellar formation. This implies that the largest episodes 
of star 
formation would leave little trace of their stellar evolution into the 
intergalactic medium leading instead to a fast metal enrichment of the
interstellar gas, as observed in high redshift quasars 
\citep[e.g.][]{1999ARA&A..37..487H,2009A&A...494L..25J}.
The hydrodynamics of star-bursting 
galaxies with a central supermassive black hole will be the subject of
a futher communication.

\acknowledgments 
We thank our anonymous referee for multiple comments and suggestions
that have largely improved the presentation of the results. 
This study has been supported by CONACYT - M\'exico, research grants
82912 and 60333, the CONACYT Mexico and the Czech Academy of Science 
research grant 2009-2010, the institutional Research Plan AVOZ10030501
of the Academy of Sciences of the Czech Republic, by the project LC06014 - 
Center for Theoretical Astrophysics of the Ministry of Education, Youth and
Sports of the Czech Republic 
and the Spanish Ministry of Science and Innovation  
under the collaboration ESTALLIDOS (grant AYA2007-67965-C03-01) and 
Consolider-Ingenio 2010 Program  grant CSD2006-00070: First Science 
with the GTC.

\bibliographystyle{aa}
\bibliography{myrefs}

\begin{thebibliography}{38}
\expandafter\ifx\csname natexlab\endcsname\relax\def\natexlab#1{#1}\fi

\bibitem[{{Baumgardt} \& {Kroupa}(2007)}]{2007MNRAS.380.1589B}
{Baumgardt}, H. \& {Kroupa}, P. 2007, \mnras, 380, 1589

\bibitem[{{Bower} {et~al.}(2004){Bower}, {Morris}, {Bacon}, {Wilman},
  {Sullivan}, {Chapman}, {Davies}, {de Zeeuw}, \&
  {Emsellem}}]{2004MNRAS.351...63B}
{Bower}, R.~G., {Morris}, S.~L., {Bacon}, R., {et~al.} 2004, \mnras, 351, 63

\bibitem[{{Chevalier}(1992)}]{1992ApJ...397L..39C}
{Chevalier}, R.~A. 1992, \apjl, 397, L39

\bibitem[{{Chevalier} \& {Clegg}(1985)}]{1985Natur.317...44C}
{Chevalier}, R.~A. \& {Clegg}, A.~W. 1985, \nat, 317, 44

\bibitem[{{Dekel} \& {Silk}(1986)}]{1986ApJ...303...39D}
{Dekel}, A. \& {Silk}, J. 1986, \apj, 303, 39

\bibitem[{{Elmegreen}(1999)}]{1999ApJ...517..103E}
{Elmegreen}, B.~G. 1999, \apj, 517, 103

\bibitem[{{Ferreras} {et~al.}(2002){Ferreras}, {Scannapieco}, \&
  {Silk}}]{2002ApJ...579..247F}
{Ferreras}, I., {Scannapieco}, E., \& {Silk}, J. 2002, \apj, 579, 247

\bibitem[{{Friaca} \& {Terlevich}(1998)}]{1998MNRAS.298..399F}
{Friaca}, A.~C.~S. \& {Terlevich}, R.~J. 1998, \mnras, 298, 399

\bibitem[{{Geyer} \& {Burkert}(2001)}]{2001MNRAS.323..988G}
{Geyer}, M.~P. \& {Burkert}, A. 2001, \mnras, 323, 988

\bibitem[{{Greve} {et~al.}(2005){Greve}, {Bertoldi}, {Smail}, {Neri},
  {Chapman}, {Blain}, {Ivison}, {Genzel}, {Omont}, {Cox}, {Tacconi}, \&
  {Kneib}}]{2005MNRAS.359.1165G}
{Greve}, T.~R., {Bertoldi}, F., {Smail}, I., {et~al.} 2005, \mnras, 359, 1165

\bibitem[{{Hamann} \& {Ferland}(1999)}]{1999ARA&A..37..487H}
{Hamann}, F. \& {Ferland}, G. 1999, \araa, 37, 487

\bibitem[{{Heckman} {et~al.}(1990){Heckman}, {Armus}, \&
  {Miley}}]{1990ApJS...74..833H}
{Heckman}, T.~M., {Armus}, L., \& {Miley}, G.~K. 1990, \apjs, 74, 833

\bibitem[{{Hughes} {et~al.}(1998){Hughes}, {Serjeant}, {Dunlop},
  {Rowan-Robinson}, {Blain}, {Mann}, {Ivison}, {Peacock}, {Efstathiou}, {Gear},
  {Oliver}, {Lawrence}, {Longair}, {Goldschmidt}, \&
  {Jenness}}]{1998Natur.394..241H}
{Hughes}, D.~H., {Serjeant}, S., {Dunlop}, J., {et~al.} 1998, \nat, 394, 241

\bibitem[{{Juarez} {et~al.}(2009){Juarez}, {Maiolino}, {Mujica}, {Pedani},
  {Marinoni}, {Nagao}, {Marconi}, \& {Oliva}}]{2009A&A...494L..25J}
{Juarez}, Y., {Maiolino}, R., {Mujica}, R., {et~al.} 2009, \aap, 494, L25

\bibitem[{{Leitherer} \& {Heckman}(1995)}]{1995ApJS...96....9L}
{Leitherer}, C. \& {Heckman}, T.~M. 1995, \apjs, 96, 9

\bibitem[{{Leitherer} {et~al.}(1999){Leitherer}, {Schaerer}, {Goldader},
  {Gonz{\'a}lez Delgado}, {Robert}, {Kune}, {de Mello}, {Devost}, \&
  {Heckman}}]{1999ApJS..123....3L}
{Leitherer}, C., {Schaerer}, D., {Goldader}, J.~D., {et~al.} 1999, \apjs, 123,
  3

\bibitem[{{Melioli} \& {de Gouveia dal Pino}(2006)}]{2006A&A...445L..23M}
{Melioli}, C. \& {de Gouveia dal Pino}, E.~M. 2006, \aap, 445, L23

\bibitem[{{Nesvadba} {et~al.}(2007){Nesvadba}, {Lehnert}, {Genzel},
  {Eisenhauer}, {Baker}, {Seitz}, {Davies}, {Lutz}, {Tacconi}, {Tecza},
  {Bender}, \& {Abuter}}]{2007ApJ...657..725N}
{Nesvadba}, N.~P.~H., {Lehnert}, M.~D., {Genzel}, R., {et~al.} 2007, \apj, 657,
  725

\bibitem[{{Plewa}(1995)}]{1995MNRAS.275..143P}
{Plewa}, T. 1995, \mnras, 275, 143

\bibitem[{{Scannapieco} {et~al.}(2002){Scannapieco}, {Ferrara}, \&
  {Madau}}]{2002ApJ...574..590S}
{Scannapieco}, E., {Ferrara}, A., \& {Madau}, P. 2002, \apj, 574, 590

\bibitem[{{Schinnerer} {et~al.}(2008){Schinnerer}, {B{\"o}ker}, {Meier}, \&
  {Calzetti}}]{2008ApJ...684L..21S}
{Schinnerer}, E., {B{\"o}ker}, T., {Meier}, D.~S., \& {Calzetti}, D. 2008,
  \apjl, 684, L21

\bibitem[{{Silich} {et~al.}(2005){Silich}, {Tenorio-Tagle}, \&
  {A{\~n}orve-Zeferino}}]{2005ApJ...635.1116S}
{Silich}, S., {Tenorio-Tagle}, G., \& {A{\~n}orve-Zeferino}, G.~A. 2005, \apj,
  635, 1116

\bibitem[{{Silich} {et~al.}(2008){Silich}, {Tenorio-Tagle}, \&
  {Hueyotl-Zahuantitla}}]{2008ApJ...686..172S}
{Silich}, S., {Tenorio-Tagle}, G., \& {Hueyotl-Zahuantitla}, F. 2008, \apj,
  686, 172

\bibitem[{{Silich} {et~al.}(2003){Silich}, {Tenorio-Tagle}, \&
  {Mu{\~n}oz-Tu{\~n}{\'o}n}}]{2003ApJ...590..791S}
{Silich}, S., {Tenorio-Tagle}, G., \& {Mu{\~n}oz-Tu{\~n}{\'o}n}, C. 2003, \apj,
  590, 791

\bibitem[{{Silich} {et~al.}(2004){Silich}, {Tenorio-Tagle}, \&
  {Rodr{\'{\i}}guez-Gonz{\'a}lez}}]{2004ApJ...610..226S}
{Silich}, S., {Tenorio-Tagle}, G., \& {Rodr{\'{\i}}guez-Gonz{\'a}lez}, A. 2004,
  \apj, 610, 226

\bibitem[{{Stone} \& {Norman}(1992)}]{1992ApJS...80..753S}
{Stone}, J.~M. \& {Norman}, M.~L. 1992, \apjs, 80, 753

\bibitem[{{Strickland} \& {Heckman}(2009)}]{2009ApJ...697.2030S}
{Strickland}, D.~K. \& {Heckman}, T.~M. 2009, \apj, 697, 2030

\bibitem[{{Strickland} \& {Stevens}(2000)}]{2000MNRAS.314..511S}
{Strickland}, D.~K. \& {Stevens}, I.~R. 2000, \mnras, 314, 511

\bibitem[{{Swinbank} {et~al.}(2006){Swinbank}, {Chapman}, {Smail}, {Lindner},
  {Borys}, {Blain}, {Ivison}, \& {Lewis}}]{2006MNRAS.371..465S}
{Swinbank}, A.~M., {Chapman}, S.~C., {Smail}, I., {et~al.} 2006, \mnras, 371,
  465

\bibitem[{{Tacconi} {et~al.}(2008){Tacconi}, {Genzel}, {Smail}, {Neri},
  {Chapman}, {Ivison}, {Blain}, {Cox}, {Omont}, {Bertoldi}, {Greve},
  {F{\"o}rster Schreiber}, {Genel}, {Lutz}, {Swinbank}, {Shapley}, {Erb},
  {Cimatti}, {Daddi}, \& {Baker}}]{2008ApJ...680..246T}
{Tacconi}, L.~J., {Genzel}, R., {Smail}, I., {et~al.} 2008, \apj, 680, 246

\bibitem[{{Tacconi} {et~al.}(2006){Tacconi}, {Neri}, {Chapman}, {Genzel},
  {Smail}, {Ivison}, {Bertoldi}, {Blain}, {Cox}, {Greve}, \&
  {Omont}}]{2006ApJ...640..228T}
{Tacconi}, L.~J., {Neri}, R., {Chapman}, S.~C., {et~al.} 2006, \apj, 640, 228

\bibitem[{{Tenorio-Tagle} \& {Bodenheimer}(1988)}]{1988ARA&A..26..145T}
{Tenorio-Tagle}, G. \& {Bodenheimer}, P. 1988, \araa, 26, 145

\bibitem[{{Tenorio-Tagle} {et~al.}(2006){Tenorio-Tagle},
  {Mu{\~n}oz-Tu{\~n}{\'o}n}, {P{\'e}rez}, {Silich}, \&
  {Telles}}]{2006ApJ...643..186T}
{Tenorio-Tagle}, G., {Mu{\~n}oz-Tu{\~n}{\'o}n}, C., {P{\'e}rez}, E., {Silich},
  S., \& {Telles}, E. 2006, \apj, 643, 186

\bibitem[{{Tenorio-Tagle} {et~al.}(2003){Tenorio-Tagle}, {Silich}, \&
  {Mu{\~n}oz-Tu{\~n}{\'o}n}}]{2003ApJ...597..279T}
{Tenorio-Tagle}, G., {Silich}, S., \& {Mu{\~n}oz-Tu{\~n}{\'o}n}, C. 2003, \apj,
  597, 279

\bibitem[{{Tenorio-Tagle} {et~al.}(2005){Tenorio-Tagle}, {Silich},
  {Rodr{\'{\i}}guez-Gonz{\'a}lez}, \&
  {Mu{\~n}oz-Tu{\~n}{\'o}n}}]{2005ApJ...628L..13T}
{Tenorio-Tagle}, G., {Silich}, S., {Rodr{\'{\i}}guez-Gonz{\'a}lez}, A., \&
  {Mu{\~n}oz-Tu{\~n}{\'o}n}, C. 2005, \apjl, 628, L13

\bibitem[{{Tenorio-Tagle} {et~al.}(2007){Tenorio-Tagle}, {W{\"u}nsch},
  {Silich}, \& {Palou{\v s}}}]{2007ApJ...658.1196T}
{Tenorio-Tagle}, G., {W{\"u}nsch}, R., {Silich}, S., \& {Palou{\v s}}, J. 2007,
  \apj, 658, 1196

\bibitem[{{Veilleux} {et~al.}(2005){Veilleux}, {Cecil}, \&
  {Bland-Hawthorn}}]{2005ARA&A..43..769V}
{Veilleux}, S., {Cecil}, G., \& {Bland-Hawthorn}, J. 2005, \araa, 43, 769

\bibitem[{{W{\"u}nsch} {et~al.}(2008){W{\"u}nsch}, {Tenorio-Tagle}, {Palou{\v
  s}}, \& {Silich}}]{2008ApJ...683..683W}
{W{\"u}nsch}, R., {Tenorio-Tagle}, G., {Palou{\v s}}, J., \& {Silich}, S. 2008,
  \apj, 683, 683

\end{thebibliography}

\end{document}